\documentclass[10pt,conference,final]{IEEEtran}
\pdfoutput=1

\def\BibTeX{{\rm B\kern-.05em{\sc i\kern-.025em b}\kern-.08emT\kern-.1667em\lower.7ex\hbox{E}\kern-.125emX}}

\newcommand\copyrighttext{%
  \footnotesize \textcopyright 2019 IEEE. Personal use of this material is permitted. Permission from IEEE must be obtained for all other uses, in any current or future media, including reprinting/republishing this material for advertising or promotional purposes, creating new collective works, for resale or redistribution to servers or lists, or reuse of any copyrighted component of this work in other works.
  DOI: TBD}

\newcommand\copyrightnotice{%
\begin{tikzpicture}[remember picture,overlay]
\node[anchor=south,yshift=10pt] at (current page.south) {\fbox{\parbox{\dimexpr\textwidth-\fboxsep-\fboxrule\relax}{\copyrighttext}}};
\end{tikzpicture}%
}

\usepackage[utf8]{inputenc}
\usepackage[T1]{fontenc}
\usepackage{microtype}

\usepackage{amssymb}
\usepackage{amsmath}

\usepackage{graphicx}
\graphicspath{{fig/}}

\usepackage[table]{xcolor}
\usepackage{multicol}
\usepackage{multirow}
\usepackage{tabularx}

\newcolumntype{C}{>{\centering\arraybackslash}X}
\newcolumntype{s}{>{\centering}m{0.1\columnwidth}}

\usepackage{listings}
\usepackage{syntax}
\usepackage{algorithm}
\usepackage[noend]{algpseudocode}

\usepackage{subcaption}
\usepackage{mdwlist}
\usepackage{tablefootnote}

\usepackage{todonotes}

\usepackage[all]{nowidow}
\usepackage{balance}

\usepackage{url}

\lstdefinestyle{basic}{
  extendedchars     = true,
  inputencoding     = utf8,
  basicstyle        = {\ttfamily \small},
  keywordstyle      = {\rmfamily \bfseries},
  commentstyle      = {\rmfamily \itshape},
  tabsize           = 2,
  flexiblecolumns   = false,
  frame             = single,
  showstringspaces  = false,
  breaklines        = true,
}

\lstdefinelanguage{Kotlin}{
  keywords = {
    package, as?, typealias,
    this, super, val, var,
    fun, for, null, true,
    false, throw,
    return, break, continue, object,
    if, try, else, while,
    do, when, class, interface,
    enum, companion, override, public,
    private, get, set, import,
    abstract, vararg, expect, actual,
    where, suspend, data, internal,
    dynamic, final, by
  },
  keywordstyle = {\bfseries},
  ndkeywords = {
    @Deprecated, @JvmName, @JvmStatic, @JvmOverloads,
    @JvmField, @JvmSynthetic, Iterable, Int,
    Long, Integer, Short, Byte,
    Float, Double, String, Runnable,
    Array
  },
  ndkeywordstyle = {\bfseries},
  emph = {
    println, return@, forEach, map,
    mapNotNull, first, filter, firstOrNull,
    lazy, delegate
  },
  emphstyle       = {},
  identifierstyle = \color{black},
  sensitive       = true,
  commentstyle    = {\color{gray}\ttfamily},
  comment         = [l]{//},
  morecomment     = [s]{/*}{*/},
  stringstyle     = {\ttfamily},
  morestring      = [b]",
  morestring      = [s]{"""*}{*"""},
}

\lstdefinestyle{src-cpp}{
  style    = basic,
  language = C++,
}

\lstdefinestyle{pascal}{
	style    = basic,
	language = Pascal,
}

\lstdefinestyle{msg}{
    style    = basic,
	language = Kotlin,
}

\usepackage{epstopdf}

\begin{document}

\title{ReduKtor: How We Stopped Worrying About Bugs in Kotlin Compiler}

\author{
\IEEEauthorblockN{Daniil Stepanov, Marat Akhin, Mikhail Belyaev}
\IEEEauthorblockA{\textit{Saint Petersburg Polytechnic University}}
\IEEEauthorblockA{\textit{JetBrains Research}\\
Saint Petersburg, Russia\\
\texttt{\{stepanov,akhin,belyaev\}@kspt.icc.spbstu.ru}}
}

\maketitle
\copyrightnotice

\begin{abstract}
Bug localization is well-known to be a difficult problem in software engineering, and specifically in compiler development, where it is beneficial to reduce the input program to a minimal reproducing example; this technique is more commonly known as delta debugging.
What additionally contributes to the problem is that every new programming language has its own unique quirks and foibles, making it near impossible to reuse existing tools and approaches with full efficiency.
In this experience paper we tackle the delta debugging problem w.r.t. Kotlin, a relatively new programming language from JetBrains.
Our approach is based on a novel combination of program slicing, hierarchical delta debugging and Kotlin-specific transformations, which are synergistic to each other.
We implemented it in a prototype called ReduKtor and did extensive evaluation on both synthetic and real Kotlin programs; we also compared its performance with classic delta debugging techniques.
The evaluation results support the practical usability of our approach to Kotlin delta debugging and also shows the importance of using both language-agnostic and language-specific techniques to achieve best reduction efficiency and performance.
\end{abstract}

\begin{IEEEkeywords}
program fuzzing, delta debugging, program slicing, input reduction, compiler testing
\end{IEEEkeywords}

\section{Introduction}
\label{sec:introduction}

In the recent years software has been becoming more and more complex, with the subsequent rise in difficulty of debugging.
Despite all the latest advances in the field of software quality assurance, such as testing, static analysis and verification, finding the root cause of a bug still takes a lot of time and effort from a developer.
One of the main reasons for that is the inherent difficulty of bug localization --- figuring out which part of the program or \emph{program inputs} causes the bug to manifest itself.
In compilers, for example, reducing the input file to a minimal reproducing example is a very important step in bug investigation, which is often done manually.

Automatic reduction of input files which cause a compiler error greatly simplifies the debugging process, as it removes irrelevant details, allowing the developer to better understand and focus on what triggers the bug.
Several techniques such as delta debugging~\cite{zeller2002simplifying}, program slicing~\cite{weiser1981program} and their various improvements~\cite{misherghi2006hdd,herfert2017automatically,binkley2014orbs} attempt to deal with this problem.
These methods are language-agnostic, which is their clear-cut advantage; however, in practice they often are hard to employ efficiently for real-world cases, with complex language-specific interdependencies and features.
Using stand-alone language-oriented transformations~\cite{sterling2007automated} or incorporating them into other reduction techniques~\cite{regehr2012test} has been shown to perform best.

This work aims at creating an automatic input reduction tool for the Kotlin%
\footnote{\url{https://kotlinlang.org/}}
compiler.
Our approach is based on a novel combination of program slicing, hierarchical delta debugging and Kotlin-specific transformations, which help each other in finding and pruning irrelevant parts of input Kotlin files.
As real-world Kotlin projects often consist of multiple files, we also support simultaneous reduction of several input files.

We have implemented our approach in a prototype tool called ReduKtor and tested it extensively on several types of input.
For the first part of the evaluation, we applied ReduKtor to a number of input files generated by Kotlin compiler fuzzer~\cite{fuzzer2017}.
For the second part of the evaluation, a number of real-world projects were injected with various bugs, which trigger Kotlin compiler errors, and processed with ReduKtor.
We also compared ReduKtor with classic input reduction techniques, such as slicing and hierarchical delta debugging~(HDD).
Our results show ReduKtor to outperform other approaches on the size and complexity of the resulting inputs, and also support the need for hybrid reduction approaches.

The rest of the paper is organized as follows.
We introduce the basics of various input reduction techniques and approaches in section~\ref{sec:basics}.
In section~\ref{sec:kotlin-reduction} we explain our approach in more detail, how it relates to Kotlin features and what challenges we had to tackle.
We talk about the implementation in section~\ref{sec:implementation}; evaluation setup and results are discussed in section~\ref{sec:evaluation}.
We overview related works in section~\ref{sec:related-work}, make conclusion and identify possible future work in section~\ref{sec:conclusion}.

\section{Input reduction 101}
\label{sec:basics}

As we established earlier, input data triggering a program error often contains a lot of irrelevant information; for example, listing~\ref{lst:when-example} shows an example Kotlin program, which crashes the compiler.
Despite the program being more than 50 lines, the actual error is triggered by the single expression \lstinline|(when{})|.
Of course, in this case the reduction could be done manually, however, such an approach does not scale well for bigger and more intricate inputs.
That is why a lot of effort has been put into automating the process of input reduction.
Let us briefly review the different techniques applicable to compiler input reduction, i.e., reduction of program source code.

\begin{figure}[t]
\begin{lstlisting}[language=Kotlin]
fun box(): String {
    return (when {
        (if ("OK") == ("OK")) {}
        // ... 50 more lines
    })
}
\end{lstlisting}
\caption{Compiler-crashing \texttt{(when\{\})} example}
\label{lst:when-example}
\end{figure}

Slicing was historically among the first techniques to attempt simplification of a given program w.r.t. number of criteria, which describe interesting program properties; it was coined by Weiser in~\cite{weiser1981program}.
A slicing criterion is often given as a pair ``program statement~-- set of variables'' and is used to create a program slice: a reduced program containing only elements related to the criterion, e.g., elements which influence the criterion variables at the given statement.
By definition, for slicing to be applicable to an input reduction problem, we must have a sound slicing criterion, a setup we do not always have.

Delta debugging is an approach free from such limitation, proposed by Zeller and Hildebrandt in~\cite{zeller2002simplifying}.
It views an input sample $T$ of a program $P$ as a combination of individual components, which can be minimized to $T_{min}$, such that:
\begin{itemize}
\item $T_{min}$ causes the same failure as $T$
\item if you remove any component of $T_{min}$, the error does not reproduce on $P$
\end{itemize}
The minimization is done via a variation of binary search; let program $P$ and input $T$ be given, such that $P(T)$ crashes the program.
We find $T_{min} = DD(T)$ using the following procedure: the initial input sample $T$ is divided into $n$ parts $T = \Delta_1 \cup \Delta_2 \cup \ldots \cup \Delta_n$, which are examined in order to understand if they could be removed from the input.
If needed, we increase $n$ and repeat the procedure.
This delta debugging algorithm from~\cite{zeller2002simplifying} is shown in figure~\ref{fig:ddmin}.
\begin{figure}[b]
$DD(T) = DD_2(T, 2)$ \\
\[
DD_2(T, n) =
  \begin{cases}
    DD_2(\Delta_i, 2)             & \quad \text{if } P(\Delta_i) \text{ fails} \\
    DD_2(\nabla_i, max(n - 1, 2)) & \quad \text{if } P(\nabla_i) \text{ fails} \\
    DD_2(T, min(|T|, 2n))         & \quad \text{if } n < |T| \\
    T                             & \quad \text{otherwise}
  \end{cases}
\]
\\
$\nabla_i = T - \Delta_i$ \\
$T = \Delta_1 \cup \Delta_2 \cup \ldots \cup \Delta_n, \forall i \neq j : \Delta_i \cap \Delta_j = \emptyset$
\caption{Delta debugging algorithm}
\label{fig:ddmin}
\end{figure}

The principal input agnosticism of delta debugging is its main pro and con; on one hand, it can be applied to any kind of input data, on the other hand, it does not take into account the inherent structure of the input and usually performs poorly on complex data.
In 2006, Mishergi and Su introduced hierarchical delta debugging~\cite{misherghi2006hdd}, an extension of classic delta debugging, which focuses on applying delta debugging to hierarchical inputs~(e.g., abstract syntax trees, XML documents, HTML pages), representable as trees.
Their algorithm works as shown in figure~\ref{fig:hdd}; it independently analyzes every level of the input tree in a top-down fashion.
First, it collects all tree nodes at the current level; second, these nodes are reduced using the classic delta debugging, creating a minimal node configuration.
This configuration is then applied to the tree, following which we proceed to the next tree level.
\newcommand{\tree}{\mathit{tree}}
\newcommand{\level}{\mathit{level}}
\newcommand{\nodes}{\mathit{nodes}}
\newcommand{\minConfig}{\mathit{minConfig}}
\begin{figure}[b]
\setlength{\leftskip}{0cm}
\textbf {INPUT:} hierarchical input data $\tree$ \\
\textbf {RESULT:} $\tree$ is reduced
\begin{algorithmic}[1]
\State $\level \Leftarrow 0 $
\State $\nodes \Leftarrow \text{getNodes}(\tree, \level)$
\While {$\nodes \neq \emptyset$}
  \State $\minConfig \Leftarrow \text{DD}(\nodes)$
  \State $\text{removeNodes}(\tree, \level, \minConfig)$
  \State $\level \Leftarrow \level + 1$
  \State $\nodes \Leftarrow \text{getNodes}(\tree, \level) $
\EndWhile
\end{algorithmic}
\caption{Hierarchical delta debugging}
\label{fig:hdd}
\end{figure}

While~HDD does work better than classic delta debugging on structured data and can achieve good reduction on any input, for many practical examples its performance is subpar, as it has no knowledge about the finer structure of the input.
For programming languages, for example, it does not take into account control or data dependencies, and has to deduce them implicitly, via its level-by-level tree reduction.
To speed up this process and improve on its quality, many practical delta debugging tools include custom language-specific transformations~\cite{sterling2007automated,regehr2012test,jsDelta}.
For example, C-Reduce attempts to replace arbitrary function arguments with compatible constants and performs inlining of small functions.

For Kotlin, we decided to combine these approaches in such a way that individual techniques support each other.
Let us describe our approach in more detail.

\section{Reduction for Kotlin compiler}
\label{sec:kotlin-reduction}

\begin{figure}[tb]
\centering
\includegraphics[width=\linewidth]{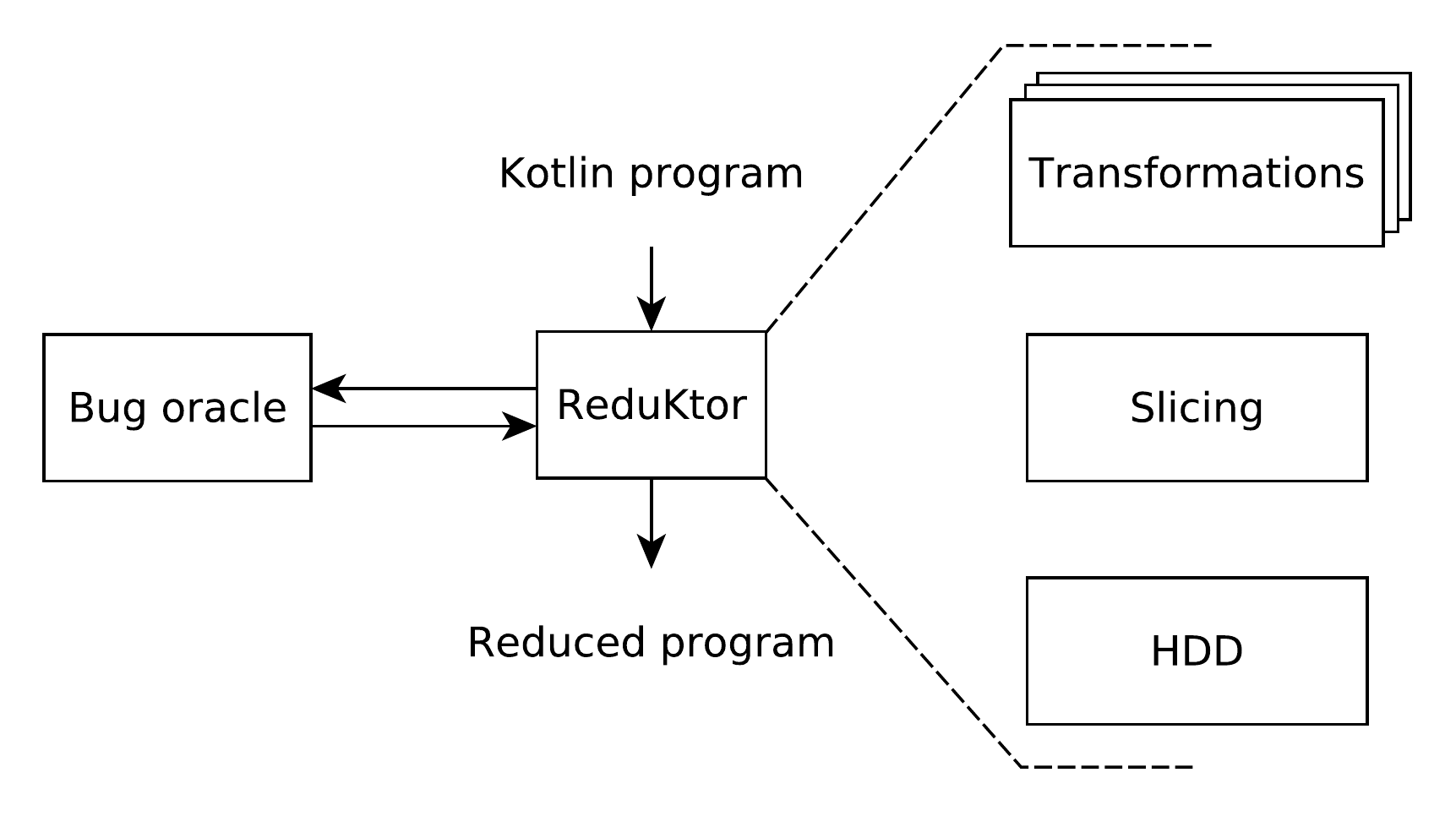}
\caption{Overview of our approach}
\label{fig:reduktor-scheme}
\end{figure}

Our approach to input reduction for Kotlin compiler~(see figure~\ref{fig:reduktor-scheme}), similarly to many other approaches, consists of a number of independent steps, each taking a set of Kotlin files as an input and reducing them via some kind of transformation.
Even before we discuss these steps in more detail, we need to explore two things: transformation soundness and their order.

\subsection{Transformation soundness}
\label{sec:soundness}

A transformation is \emph{sound} w.r.t. input reduction problem, if it preserves the original error; if a program $P$ fails on input $T$, it should fail with the same error on the transformed input $T^{\prime}$.
However, for compiler errors it is often very hard to define what ``same'' means.
For example, an error message may contain parts of generated bytecode, which may be different for files before and after transformation, but relate to the same cause.

There are 2 ways to solve this problem.
The first one assumes error messages have a specific format, which can be automatically parsed and compared.
A number of errors from Kotlin compiler satisfy this condition; an example of such error is shown in listing~\ref{lst:error-msg-example}.
As one can see from the example, the type and location of error are easily identifiable.
For such errors we have developed a number of parsers, which extract relevant information used for error comparison.

If error messages do not have a structure, we need to compare them directly, i.e., find their differences.
A classic approach is to view this problem as a string difference one, solvable via longest common subsequence~(LCS) algorithm, such as Myers'~\cite{myers1986ano}.
To avoid reinventing the wheel, we use Diff     Match Patch~\cite{diff-match-patch}, a highly-optimized library from Google for text synchronization.
The Diff component compares two strings and outputs a list of their differences, which we can use to estimate the similarity of original strings as $\sum |\mathit{diff}| / \sum |\mathit{total}|$.
The closer this coefficient is to zero, the more similar the strings are.
To be considered the same, error messages should have similarity lower than a set threshold, which is configurable.

Our approach first attempts to compare two error messages using one of the template parsers; if they fail, it falls back to a direct difference comparison.
This allows our approach to more aggressively reduce input for a large class of errors, which have comprehensive error messages.

\begin{figure}[tb]
\begin{lstlisting}[style=msg]
Error:Kotlin: [Internal Error] org.jetbrains.kotlin.codegen.CompilationException: Back-end (JVM) Internal error: wrong code generated
org.jetbrains.kotlin.codegen.CompilationException Back-end (JVM) Internal error: Couldn't transform method node:
test ()V:
   L0
    LINENUMBER 2 L0
   L1
    POP
   L2
...
Cause: AFTER mandatory stack transformations: incorrect bytecode
Element is unknown The root cause was thrown at: MethodVerifier.kt:28
File being compiled at position: (1,1) in Main.kt
...
\end{lstlisting}
\caption{An example of a Kotlin compiler error message with well-defined format}
\label{lst:error-msg-example}
\end{figure}

\subsection{Transformation order}
\label{sec:order}

All input reduction steps are independent of each other, however, even for a fixed set of transformations, the final result may be affected by the \emph{order}, in which these transformations are applied.
More so, the overall performance also depends on the transformation order, as the individual performance of many transformations is greatly related to the input size.
For this work, we decided to apply transformations in the order of their ``reductiveness'' based on the preliminary evaluation, from coarse-grained steps, which can remove whole input files, to more fine-grained ones, which may work on a subexpression level.
It would be interesting to see, if it is possible to reliably learn a quazi-optimal transformation order for a specific class of inputs via machine learning or genetic engineering.
We hope to explore this opportunity in our future work.

\vskip 1em

The resulting pipeline is as follows.
\begin{itemize}
	\item Project-level simplifications
	\item Slicing
	\item Text based transformations
	\item Syntax tree based transformations
	\item Hierarchical delta debugging
\end{itemize}
Every transformation is applied in order until convergence, i.e., until its input stops decreasing in size.
We also check if the transformation preserves the error by comparing before and after error messages.
If the error stops reproducing, we rollback the input and continue with the subsequent transformations.

\subsection{Reduction criterion}
\label{sec:rc}

As mentioned before, our approach takes as an input a set of Kotlin files with its respective error, which allows to support reduction of not only individual files, but also whole projects.
However, some transformations need a starting point, e.g., slicing needs a criterion from which to do its work.
When the error message contains information about the error location, we extract it as described in~\ref{sec:soundness} and use from here on.
If the information is missing, we ask the developer to provide a file, presumably containing the error.
From here on out, we will call this starting point a \emph{reduction criterion}~(RC).

{\em Note:} despite our fallback strategy appearing as unsound~(relying on manual developer input may seem quite unsound), the approach established in section~\ref{sec:soundness} helps us to ensure soundness.
If the developer specifies an incorrect file for the~RC, they will get a suboptimal reduction, but the original error will still be preserved.
In case you are interested in how often the fallback~RC have been actually used, we kindly forward you to the evaluation, specifically section~\ref{sec:evaluation-soundness}.

\subsection{Project-level simplifications}
\label{sec:project-simplifications}

The first transformation we use performs different project-level simplifications.
Their main purpose is to prune away parts of the project which are irrelevant to the error.
Modern programs have complex internal dependencies, greatly complicating reducing individual files.
To solve this problem, we need to remove these dependencies.

In Kotlin dependencies of a file are usually defined via import lists: statements specifying which other parts of the project are needed in the current file.
Another way of specifying dependencies is fully qualified names~(FQN), when a program component is referenced by its complete name.
These elements form the \emph{dependency tree}, which may be used to guide the simplification process.

\begin{figure}[tb]
\begin{subfigure}[t]{\linewidth}
\caption{Example of import dependencies}
\label{list:dep-example-a}
\begin{lstlisting}[language=Kotlin]
// File A.kt
class A() {
    fun funWithBug() {
        // ...
    }
    
    fun f() {
        // ...
    }
}

// File B.kt
class B() {
    val a = A()
    
    fun f() = a.f()
}

// File C.kt
class C() {
    val a = A()
}

// File D.kt
class D() {
    val b = B()
    
    fun f() = b.f()
}
\end{lstlisting}
\end{subfigure}

\begin{subfigure}[t]{\linewidth}
\caption{Built dependency tree}
\label{lst:dep-example-b}
\centering
\includegraphics[width=0.38\linewidth]{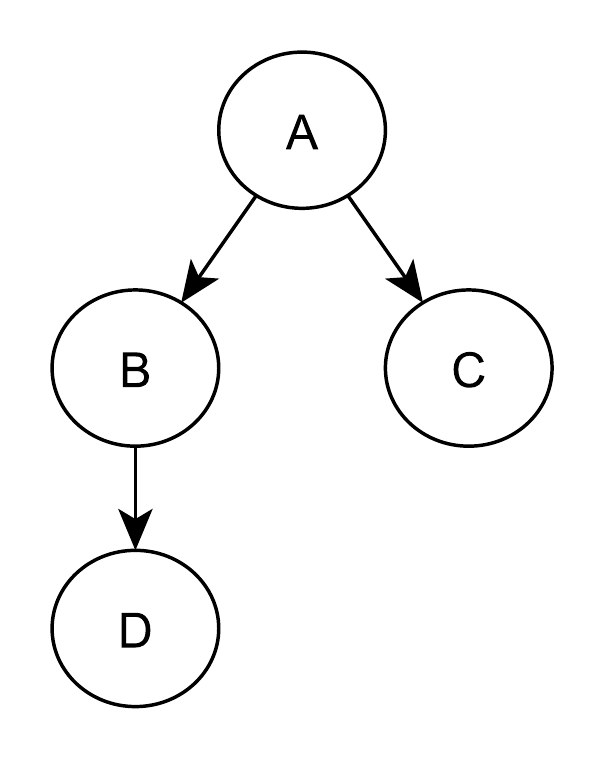}
\end{subfigure}

\caption{Example of project dependencies problem}
\label{lst:dep-example}
\end{figure}

Figure~\ref{list:dep-example-a} shows an example with import dependencies. In order to reduce class A and leave only the relevant function with the bug, we must reduce classes B, C and D in the correct order. To do this we use the dependency tree, as in figure~\ref{lst:dep-example-b}.

To build this tree, we recursively collect the imports and FQNs, starting from the reduction criterion.
All files not in the dependency tree are removed from the project.
After that, we traverse the dependency tree in the bottom up order, applying to each file a subset of transformations, aimed specifically at simplifying the project dependencies.
As all files except one do not contain the RC, all transformations must not depend on a given starting point.
The transformations used are listed below, we describe them in more detail in their respective sections.
\begin{itemize}
\item All text based transformations
\item Syntax tree based transformations related to pruning unused program parts
\begin{itemize}
    \item Removing of unused components
    \item Simplifying interdependent components
\end{itemize}
\end{itemize}

\subsection{Slicing}
\label{sec:slicing}

As mentioned in section~\ref{sec:basics}, slicing is one of the most basic input reduction techniques.
It creates a program slice, free of unneeded parts w.r.t. slicing criterion; in our case, we use the RC as the slicing criterion.
There are several different types of slicing varying in their performance and complexity.
For our purposes we decided to implement a static backward slicing~\cite{2002slicing} over the syntax trees, which is applied on the following levels in their respective order.
\begin{itemize}
\item Intraprocedural level
\item Function level
\item Class level
\end{itemize}

The algorithm for intraprocedural slicing is presented in figure~\ref{alg:interprocedural-slicing}.
\newcommand{\targetLine}{\mathit{targetLine}}%
\newcommand{\curLine}{\mathit{curLine}}%
\newcommand{\lastLine}{\text{getLastLine}(\fun)}%
\newcommand{\firstLine}{\text{getFirstLine}(\fun)}%
\newcommand{\fun}{\mathit{fun}}%
\newcommand{\deps}{\mathit{deps}}%
\begin{figure}[tb]
\setlength{\leftskip}{0cm}
\textbf{INPUT:} slicing criterion $\targetLine$ \\
\textbf{INPUT:} function $\fun$ \\
\textbf{OUTPUT:} reduced function $\fun$ \\
\begin{algorithmic}[1]
\State $\curLine \leftarrow \lastLine$
\While{$\curLine \neq targetLine$} 
	\State $\text{deleteLineIfSound}(\fun, \curLine)$
	\State $\curLine \leftarrow \text{getPrevLine}(\fun, \curLine)$
\EndWhile
\State $\deps \leftarrow \text{getDeps}(\curLine)$
\While{$\curLine \geq \firstLine$} 
	\If{$\text{dependsOn}(\curLine, \deps)$} 
		\State $\deps \leftarrow \deps + \text{getDeps}(\curLine)$ 
	\Else 
		\State $\text{deleteLineIfSound}(\fun, \curLine)$ 
	\EndIf
	\State $\curLine \leftarrow \text{getPrevLine}(\fun, \curLine)$
\EndWhile
\State \Return $\fun$
\end{algorithmic}
\caption{Intraprocedural slicing algorithm}
\label{alg:interprocedural-slicing}
\end{figure}
The algorithm analyzes the function in the reverse order.
All lines after the slicing criterion are removed using the $\text{deleteLineIfSound}$ function, which checks for transformation soundness, as described in sections~\ref{sec:soundness} and~\ref{sec:order}.
After that we collect the interesting variables~(aka dependencies) from the slicing criterion and continue processing the function; if a line depends on these variables, we keep it and extend the dependencies, otherwise it can be removed.
The $\text{dependsOn}$ and $\text{getDeps}$ functions can handle different Kotlin language statements and expressions~(such as if statements, loops, etc.).
In case the RC does not contain an exact line, we perform redundant slicing w.r.t. every line as the possible slicing criterion and pick the best~(shortest) resulting file.

\newcommand{\targetFun}{\mathit{targetFun}}%
\newcommand{\file}{\mathit{file}}%
\newcommand{\callTree}{\mathit{callTree}}%
\newcommand{\buildCallTree}{\text{buildCallTree}}%
\newcommand{\callees}{\mathit{callees}}%
\newcommand{\collectCalleesFor}{\text{collectCalleesFor}}%
\newcommand{\funsToRemove}{\mathit{funsToRemove}}%
\newcommand{\getAllFuns}{\text{getAllFuns}}%
\newcommand{\topologicalSort}{\text{topologicalSort}}%
\newcommand{\removeFunIfSound}{\text{removeFunIfSound}}%

\newcommand{\getDirectCallees}{\text{getDirectCallees}}%
\newcommand{\directCallees}{\mathit{directCallees}}%
\newcommand{\res}{\mathit{res}}%
\newcommand{\direct}{\mathit{direct}}%

Slicing at the function level works similarly to the dependency tree pruning.
Starting from the slicing criterion, we collect the function call graph tree, marking all functions called as used.
Afterwards all unused functions are removed from the file, if their deletion is sound w.r.t.~RC~(see figure~\ref{alg:fun-slicing}).
Class level slicing is performed in a parallel fashion.

\algdef{SE}[FUNCTION]{Function}{EndFunction}%
   [2]{\algorithmicfunction\ \textproc{#1}\ifthenelse{\equal{#2}{}}{}{(#2)}}%
   {\algorithmicend\ \algorithmicfunction}

\begin{figure}[tb]
\setlength{\leftskip}{0cm}
\textbf{INPUT:} slicing criterion $\targetFun$ \\
\textbf{INPUT:} file $\file$ \\
\textbf{OUTPUT:} reduced file $\file$ \\
\begin{algorithmic}[1]
\State $\callTree \leftarrow \buildCallTree(\file)$
\State $\callees \leftarrow \collectCalleesFor(\targetFun, \callTree)$
\State $\funsToRemove \leftarrow \getAllFuns(\file) - \callees$
\State $\topologicalSort(\funsToRemove)$
    \For{$\fun \in \funsToRemove$}
        \State $\removeFunIfSound(\file, \fun)$
    \EndFor
\State \Return $\file$
\\
\Function{collectCalleesFor}{$\fun, \callTree$}
    \State $\directCallees \leftarrow \getDirectCallees(\fun, \callTree)$
    \State $\res \leftarrow \directCallees$
    \For{$\direct \in \directCallees$}
        \State $\res \leftarrow \res + \collectCalleesFor(\direct, \callTree)$
    \EndFor
    \State \Return $\res$
\EndFunction

\end{algorithmic}
\caption{Function-level slicing algorithm}
\label{alg:fun-slicing}
\end{figure}

An example of how our slicing algorithm works is shown in figure~\ref{fig:slicing-example}; in the example, we attempt slicing w.r.t. line~\ref{lst:slicing-criterion}.

\begin{figure}
\begin{subfigure}[t]{\linewidth}
\caption{Original example}
\begin{lstlisting}[
  language=Kotlin,
  escapechar=|,
  numbers=right
]
class Square(private val a: Double) {
    fun getPerimeter(): Double = a * 4
    fun getSquare(): Double = a * a
}

class Triangle(private val a: Double,
               private val b: Double, 
               private val c: Double) {

    fun getPerimeter(): Double = a + b + c

    fun getSquare(): Double {
        var square = 0.0
        if (a * a + b * b == c * c) {
           |\textcolor{blue}{square = a * b / 2}||\label{lst:slicing-criterion}|
        } else {
            val p = getPerimeter() / 2
            square = Math.sqrt(p * (p - a) *
                       (p - b) * (p - c))
        }
        return square
    }
}
\end{lstlisting}
\end{subfigure}

\begin{subfigure}[t]{\linewidth}
\caption{Example after slicing}
\begin{lstlisting}[
  language=Kotlin,
  escapechar=|,
  numbers=right
]
class Triangle(private val a: Double,
               private val b: Double, 
               private val c: Double) {

    fun getSquare(): Double {
        var square = 0.0
        if (a * a + b * b == c * c) {
            |\textcolor{blue}{square = a * b / 2}|
        } else { }
        return square
    }
}
\end{lstlisting}
\end{subfigure}

\caption{Slicing algorithm: before and after example}
\label{fig:slicing-example}
\end{figure}

\subsection{Text based transformations}
\label{sec:text-transformations}

Direct transformations over the program text representation, despite their simplicity, are often the most efficient way of reducing the program. 
Examples of such transformations include removing some or all text in a string literal, changing code based on a pattern, etc.
We selected about 30 of various text transformations to be used based on the following.
\begin{itemize}
    \item Our Kotlin programming experience
    \item Transformations used in other reduction tools~(\cite{regehr2012test,WR:04})
    \item Transformations used in the Kotlin IntelliJ IDEA plugin for code simplification%
    \footnote{\url{https://www.jetbrains.com/idea/}}
\end{itemize}

As most of them are pretty straightforward to invent and implement, we skip on describing them for brevity. 
Below is an incomplete list of text based transformations included in our approach.
\begin{itemize}
    \item Removal of text inside a balanced pair of parentheses
    \item Pattern-based removal or substitution of text
    \item Replacement of string literals with empty strings
    \item Replacement of integer constants with 0
\end{itemize}
An example of these transformations is shown in figure~\ref{fig:text-transformation-example}.

\begin{figure}
\begin{subfigure}[t]{\linewidth}
\caption{Original example}
\begin{lstlisting}[language=Kotlin]
fun f() {
    var a = 815162342
    val b = a + 1
    val c = 1.1
    var d: Double
    while (a.toDouble() != c) {
        d = a * b * c
        a += 1
    }
    println("a = $a")
}
\end{lstlisting}
\end{subfigure}

\begin{subfigure}[t]{\linewidth}
\caption{Example after text based transformations}
\begin{lstlisting}[language=Kotlin]
fun f() {
    var a = 0
    val b = a + 1
    val c = 0.0
    var d: Double
    if (a.toDouble() != c) {
        d = a * c
        a++
    }
    println("")
}
\end{lstlisting}
\end{subfigure}

\caption{Text based transformations: before and after example}
\label{fig:text-transformation-example}
\end{figure}

\subsection{Syntax tree based transformations}
\label{sec:syntax-tree-transformations}

Another group of ad hoc transformations, which perform well for source code reduction, are syntax tree based transformations.
There are two main approaches to them: language-agnostic and language-specific.
Language-agnostic transformations assume only the basic tree structure and are defined in terms of generic node transformations; this allows them to successfully reduce any input representable as a tree.
On the other hand, language-specific transformations usually depend on particular tree properties for a given programming language; this limits their generality, but improves the reduction efficiency.

Similarly to text-based transformations~(section~\ref{sec:text-transformations}), we decided to use a number of Kotlin-specific transformations, derived from our Kotlin programming experience and from which transformations are used in other tools.
They can be divided into the following groups.
\begin{itemize}
    \item Expression simplification (if statements, loops, elvis operator, etc.)
    \item Removal of unneeded components (function and constructor parameters, imports, etc.)
    \item Simplification of interdependencies (removal of inherited properties and functions, replacement of function bodies with \lstinline{TODO()}, etc.)
    \item Miscellaneous (comment deletion, replacement of function return value, etc.)
\end{itemize}
Such transformations are implemented as syntax tree based ones instead of text based, because they either cannot be expressed or would create too many syntactically incorrect programs if done over text. The latter may greatly influence the overall reduction performance, as every incorrect transformation causes a rollback to the previous reduction state.

Our approach currently includes 27 Kotlin-specific syntax tree-based transformations; as describing each and every of these transformation would have taken up most of the paper, we decided to describe in detail only the most interesting ones.

\subsubsection{Simplifying elvis operator}
\label{sec:elvis}

Kotlin has an elvis operator, a succinct way of checking value for~\lstinline{null} and providing a sensible default option; \lstinline{val a = b ?: c} means ``if \lstinline{b} is not \lstinline{null}, assign the non-null value \lstinline{b} to \lstinline{a}, else assign the default value \lstinline{c} to \lstinline{a}''.
This operator can be reduced as \lstinline{val a = c}, i.e., we can drop its left-hand side.
As the type of \lstinline{c} is guaranteed to be a subtype of \lstinline{b ?: c}, this substitution is safe.

\subsubsection{Deleting function parameters}
\label{sec:function-arguments}

Function parameters often become unused as a result of other reduction steps.
To delete them efficiently, you have to perform a bona fide refactoring: modify the function itself, find all calls to the function and delete the corresponding argument.
At the same time, you have to consider function overloading and inheritance, making this transformation quite sophisticated to implement.

\subsubsection{\lstinline{TODO()} simplification}
\label{sec:todo-simplification}

Kotlin has a special \lstinline{TODO} function, which throws a \lstinline{NotImplementedError} exception when called and has a special \lstinline{Nothing} return type.
\lstinline{Nothing} is a uninhabited subtype of all Kotlin types, i.e., can be used in place of any other Kotlin expression.
The corresponding transformation attempts to replace arbitrary expressions, such as function bodies or variable initializers, with the call to \lstinline{TODO}.

\subsubsection{Inlining}
\label{sec:inlining}

In case of small functions it makes sense to inline their bodies in place of their calls, to improve readability and give other transformations additional opportunities for reduction.
The inlining threshold is configurable by the user, by default we attempt inlining for functions less than 10 lines.

\subsubsection{Simplifying if statements}
\label{sec:if-simplification}

This transformation attempts to replace the if statement with its true or false branch.
If the condition contains a type check, Kotlin performs a \emph{smart cast}%
\footnote{https://kotlinlang.org/docs/reference/typecasts.html},
a variation of flow-based typing, aimed at reducing the code boilerplate by automatically changing the compile-time type of a variable after type checks and type-check-like constructions.
For our transformation, we create a corresponding type cast~(\lstinline{v as T}) for every type check from the condition~(\lstinline{v is T}), taking into account their negation for the false branch.

\subsection{Hierarchical delta debugging}
\label{sec:hdd}

The last step in input reduction is hierarchical delta debugging~\cite{misherghi2006hdd}.
HDD is used as a finishing tool to remove redundant constructions not considered by the previous steps.
After other transformations have done their job, the input file is already significantly reduced compared to the starting point; this means HDD can be performed much faster, if we were to compare it to an HDD-only approach.

\section{Implementation}
\label{sec:implementation}

We implemented a prototype tool for Kotlin file reduction based on our approach, called ReduKtor.
In this section we would like to discuss some of the more interesting ReduKtor implementation details.

\subsection{Working with Kotlin}
\label{sec:working-with-kotlin}

Many of our reduction steps require quite an advanced understanding of the Kotlin source code, e.g., building and manipulating its abstract syntax trees~(AST).
We also need to be able to efficiently recompile the source code after each change, so that we can check if the target error is preserved, as described in section~\ref{sec:soundness}.

To achieve these goals, we made a well-considered decision to build ReduKtor on top of the Kotlin compiler using it as a library.
This allows us to use its robust source code parser, which produces Program Structure Interface~(PSI) trees%
\footnote{\url{https://www.jetbrains.org/intellij/sdk/docs/basics/architectural_overview/psi.html}},
JetBrains' traditional concrete syntax tree~(CST) implementation, supporting both text- and tree-based transformations.
By using the compiler as a library, we significantly reduced the time needed to check the error reproducibility, as Kotlin compiler has quite a long startup time, if used externally.

Unfortunately, this decision has a serious drawback: by being dependent on the very thing we are trying to debug, we may be reducing the space of supported Kotlin inputs; if the parser itself fails, ReduKtor also cannot work.
That said, in our experience we have \emph{never} encountered such a situation.

The Kotlin compiler is used in a two-stage process.
First, we use only the parser to create the PSI, which is much quicker than invoking the full compilation; if the PSI contains error nodes, the input file is syntactically incorrect and should be rejected.
Second, we perform the full compilation and analyze the error message.
This scheme also helps to save time checking for transformation soundness.

\subsection{Parallel processing}
\label{sec:parallel-processing}

The first step of the transformations~(project-level simplifications) is performed in parallel on every project file, to better utilize the modern hardware.
As described in section~\ref{sec:project-simplifications}, we process the project dependency tree in the bottom up order, and do so in parallel for independent files, i.e., viewing the dependency tree as a parallel task graph.
Additionally, we also considered running the separate reduction steps in parallel, but decided against it in the prototype, due to the complexity of how to merge the possibly interdependent results together.

\subsection{Caching}
\label{sec:caching}

Another optimization employed in ReduKtor is the caching of intermediate results.
During transformations we may encounter source code configurations, which have been already explored; this is most often encountered during~HDD.
To avoid rechecking, we cache previously checked AST configurations as their hashes together with the result.
If the current hash has already been seen, we reuse the cached result to guide the subsequent transformations.
In our experience, this significantly improves the HDD performance.

\section{Evaluation}
\label{sec:evaluation}

For the evaluation, we run ReduKtor for Kotlin compiler version 1.3.10 on two types of input.
For the first part, we used the results of Kotlin compiler fuzzer~\cite{fuzzer2017}: single files, which cause compiler crashes,~--- together with the code samples from various compiler bugs, collected from the Kotlin compiler bug tracker%
\footnote{\url{https://youtrack.jetbrains.com/issues/KT}}.
For the second part, a number of real-world projects were injected with invalid code, to test the relevance and performance of project-level simplifications.
We did attempt to find project-level inputs with compiler-crashing bugs to no avail, which is why we opted to create such inputs artificially.
The brief description of selected projects is shown in table~\ref{tab:projects}.

\begin{table*}[tb]
\center
\small{
\begin{tabular}{| c | r | r | c |}
\hline
\bf Name & \bf Lines $\times 10^3$ & \bf Tokens $\times 10^3$ & \bf Description\\
\hline
fuzzTests  & 8   & 128.8 & Kotlin fuzzer tests~(446 files)\\
\hline
bugsTests  & 1.3 & 11 & Code samples from the bug tracker~(93 files)\\
\hline
kotlinpoet & 10  & 52.4 & Library for Kotlin source code generation\\
\hline
kfg        & 3.5 & 66 & CFG builder for JVM bytecode\\
\hline
mapdb      & 2   & 14.8 & Embeddable database\\
\hline
koin       & 8   & 17.8 & Library for dependency injection\\
\hline
\end{tabular}
}
\caption{Description of the test projects}
\label{tab:projects}
\end{table*}

We collected the following statistics: reduction efficiency~(i.e., the decrease in file size) and performance~(i.e., the wall-clock time for reduction).
For every test project, we ran input reduction in the following configurations.
\begin{itemize}
\item Slicing only~(S)
\item Kotlin-specific transformations~(KST) only~(T)
\item HDD only~(D)
\item Pardis only~(P)
\item ReduKtor without transformations~(S + D)
\item ReduKtor without slicing~(T + D)
\item ReduKtor without HDD~(S + T)
\item ReduKtor~(S + T + D)
\item ReduKtor with Pardis instead of HDD~(S + T + P)
\end{itemize}
Pardis~\cite{gharachorlu2019priority} is the latest addition to the family of language-agnostic input reduction algorithms, considered to be state-of-the-art and outperform C-Reduce by a factor of 2 in reduction performance.
For the purposes of evaluation, we implemented Pardis as an optional step in ReduKtor, to see how it performs within a complex input reduction tool.

For the test bench, we used a machine with Intel Core i7-4790 3.6 GHz processor, 32 GB of memory and Intel 535 SSD storage.
The evaluation results are shown in tables~\ref{tab:res-size} and~\ref{tab:res-time}.
An example of how different modes of ReduKtor perform on an example from Kotlin fuzzer tests is shown in figure~\ref{lst:reduktor-example}.

\begin{table*}[tb]
\center
\begin{tabular}{| c | r | r | r | r | r | r | r | r | r | r | r |}
\hline
\bf \multirow{2}{*}{Project} & \multicolumn{1}{c|}{\multirow{2}{*}{\bf O(\#)}} & \multicolumn{9}{c|}{\bf R(\#)}\\
\cline{3-11}
& & \multicolumn{1}{c|}{\bf S} & \multicolumn{1}{c|}{\bf T} & \multicolumn{1}{c|}{\bf D} & \multicolumn{1}{c|}{\bf P} & \multicolumn{1}{c|}{\bf S + D} & \multicolumn{1}{c|}{\bf T + D} & \multicolumn{1}{c|}{\bf S + T} & \multicolumn{1}{c|}{\bf S + T + D} & \multicolumn{1}{c|}{\bf S + T + P}\\
\hline
fuzzTests & 128,843
& 97,269 & 39,547 & 39,546 & 43,614 & 39,587 & 22,497 & 34,181 & 22,682 & 22,623\\
\hline
bugsTests & 11,080
& 10,881 & 8,006 & 8,364 & 8,290 & 8,364 & 7,145 & 7,683 & 7,145 & 7,088\\
\hline
kotlinpoet & 8,251
& 6,595 & 605 & 462 & 525 & 462 & 32 & 644 & 32 & 32\\
\hline
kfg & 6,762
& 6,490 & 327 & 327 & 199 & 327 & 136 & 438 & 136 & 136\\
\hline
mapdb & 4,120
& 3,543 & 561 & 574 & 574 & 574 & 305 & 478 & 303 & 305\\
\hline
koin & 1,080
& 639 & 281 & 466 & 466 & 466 & 79 & 281 & 79 & 79\\
\hline
\end{tabular}
\\ \ \\ \ Column O shows the number of original tokens, column R~--- number of tokens after reduction
\caption{Reduction efficiency}
\label{tab:res-size}
\end{table*}

\begin{table*}[tb]
\center
\begin{tabular}{| c | r | r | r | r | r | r | r | r | r | r |}
\hline
\bf \multirow{2}{*}{Project} & \multicolumn{9}{c|}{\bf T(s)}\\
\cline{2-10}
& \multicolumn{1}{c|}{\bf S} & \multicolumn{1}{c|}{\bf T} & \multicolumn{1}{c|}{\bf D} & \multicolumn{1}{c|}{\bf P} & \multicolumn{1}{c|}{\bf S + D} & \multicolumn{1}{c|}{\bf T + D} & \multicolumn{1}{c|}{\bf S + T} & \multicolumn{1}{c|}{\bf S + T + D} & \multicolumn{1}{c|}{\bf S + T + P}\\
\hline
fuzzTests
& 512 & 2,023 & 4,869 & 2,009 & 4,554 & 3,215 & 1,704 & 3,018 & 2,251 \\
\hline
bugsTests
& 47 & 186 & 770 & 225 & 765 & 624 & 220 & 661 & 401\\
\hline
kotlinpoet
& 155 & 1,814 & 4,810 & 2,769 & 4,800 & 1,280 & 1,362 & 1,270 & 1,205\\
\hline
kfg
& 329 & 1,159 & 3,386 & 1,269 & 3,299 & 1,306 & 1,195 & 1,297 & 1,196\\
\hline
mapdb
& 83 & 166 & 1,282 & 335 & 758 & 431 & 169 & 428 & 267\\
\hline
koin
& 49 & 382 & 619 & 188 & 456 & 466 & 384 & 484 & 419\\
\hline
\end{tabular}
\\ \ \\ \ Column T shows the reduction wall-clock time in seconds
\caption{Reduction performance}
\label{tab:res-time}
\end{table*}

\begin{figure}[tb]

\begin{subfigure}[t]{\linewidth}
\caption{Original test case}
\begin{lstlisting}
class A() {
    fun String.test(OK: String) {}
}

fun box(): String {
    val clazz = (A)?::class.java
    val method = clazz.getDeclaredMethod("test",
        String::class.java, String::class.java)
    val parameters = method.getParameters()
    if (!method[0].isImplicit() ||
        parameters[0].isSynthetic()) {
        return "wrong modifier"
    }
    return parameters[1].name
}
\end{lstlisting}
\end{subfigure}

\begin{subfigure}[t]{\linewidth}
\caption{Test case after slicing}
\begin{lstlisting}
class A() {}

fun box(): String {
	val clazz = (A)?::class.java
	val method = clazz.getDeclaredMethod("test",
	    String::class.java, String::class.java)
	val parameters = method.getParameters()
	return parameters[1].name
}
\end{lstlisting}
\end{subfigure}

\begin{subfigure}[t]{\linewidth}
\caption{Test case after HDD}
\begin{lstlisting}
fun box(): String {
	val clazz = (A)?::class.java
	val method = clazz.getDeclaredMethod
	val parameters = method.getParameters
	return parameters.name
}
\end{lstlisting}
\end{subfigure}

\begin{subfigure}[t]{\linewidth}
\caption{Test case after full ReduKtor mode}
\begin{lstlisting}
fun box() {
	val clazz = (A)?::class.java
}
\end{lstlisting}
\end{subfigure}

\caption{Example of different reduction modes}
\label{lst:reduktor-example}
\end{figure}

\subsection{Reduction soundness}
\label{sec:evaluation-soundness}

In section~\ref{sec:soundness} we discussed two ways of ensuring reduction soundness: one based on parsing error messages in a specific format and another based on generic string comparison.
In our evaluation {\em we did not encounter a single crash}, for which our error parsers failed to extract the error type and location; this fact speaks highly of Kotlin compiler team and their error handling discipline.
This also means our evaluation compares and talks about {\em sound} reduction results.

In a scenario when ReduKtor has to fallback to generic string comparison, one has a possible problem with reduction soundness only in a large-scale evaluation, when it is not feasible to manually manage the similarity threshold.
If we are to talk about using ReduKtor in practice, there should be a human developer in the reduction feedback loop, who can check and tweak the similarity threshold as needed.

\subsection{Lessons learned}

There are several interesting insights we can extract from the evaluation results.
First, slicing impact on the input reduction is negligible: if we remove slicing from the pipeline, the reduction efficiency is almost unchanged, and the performance is decreased by 6 percent at most.
In our opinion, this means slicing is better used for subsequent bug \emph{localization}, a step performed \emph{after} the input reduction has been done.

Second, custom language-specific transformations outperform even the state-of-the-art language-agnostic technique~(Pardis) in both performance and efficiency.
They are better in reduction efficiency by up to 1.64x for 4 out of 6 project~(fuzzTests, bugsTests, mapdb, koin), and faster by up to 2.0x for 4 out of 6 projects~(bugsTests, kotlinpoet, kfg, mapdb).
This proves their importance for reduction in real-world applications.

Third, despite their standalone efficiency, the results of language-specific transformations can be significantly improved by applying language-agnostic techniques~(such as HDD or Pardis) to their results.
Adding either HDD or Pardis, together with slicing, improves the results by 1.1x to 18.9x times.
Of course, adding additional reduction steps causes a subsequent reduction in performance, by up to 3.5x times.
Besides that, from a practical standpoint, aggressive caching of intermediate results together with using compiler in a two-stage process~(see section~\ref{sec:working-with-kotlin} for further details) greatly improves performance.
For example, disabling these features slows down the reduction process for bugsTests by 12.0x times.

We believe the second and third insights to strongly support the need for hybrid input reduction approaches, to utilize the synergy between language-agnostic and language-specific techniques.
They also show we need a better strategy for comparing different input reduction tools; for example, Pardis is said to be 4 times faster than C-Reduce, however, it is a consequence of C-Reduce using HDD as its final step.

We also performed a manual overview of the reduction results and identified some inefficiencies.
In several cases, ReduKtor failed to remove all information irrelevant to the error from the target file in a multi-file project.
This happened because the project-level simplifications failed to remove all unneeded dependencies inside the project.
In our future work we plan to explore how one may better perform project-level reduction, by either extending the number of transformations used or selecting better project-level transformations.

As mentioned in section~\ref{sec:order}, the reduction process consists of many stages, the order of which can affect both the reduction time and its result.
We conducted additional experiments~(not presented here for the sake of brevity), which have shown the selected order to outperform alternative orders.
For example, if you implement the same transformations in {\em reverse} order, the reduction slows down by as much as 40\% on fuzzTests, and by 20\% on average.

\subsection{Real-world adoption}

Unfortunately, ReduKtor is only a research prototype, and we do not have any solid practical adoption yet; besides the evaluation, we only used it internally~(in our research group) for several quirky bugs triggered by our code in the Kotlin compiler.
One of the main reasons for that is, as discussed in section~\ref{sec:working-with-kotlin}, to speed up reduction we use the compiler as a library, via a non-stable internal API.
As this API changes between versions, one cannot currently use ReduKtor as an off-the-shelf solution for their particular compiler version and/or development environment.

We recently contacted the Kotlin compiler team and proposed them to incorporate ReduKtor in their workflow.
Their feedback was positive; with their cooperation we should be able to migrate ReduKtor to a new stable API, to better support different compiler versions.
We also further confirmed the need for project-level simplification in real use cases, as currently developers spend a lot of their time manually reducing complex project setups to create minimal reproducing examples.
We hope to explore how we can improve on reduction for multi-file projects in our future work.

\section{Related work}
\label{sec:related-work}

As mentioned before, input reduction is a popular research area with a lot of practically applicable tools.
At the moment, the most developed tool for delta debugging is C-Reduce~\cite{regehr2012test}, initially created for reduction of C/C\texttt{++} compiler tests, generated by Csmith~\cite{yang2011finding}.
These tests, being randomly generated, usually contain a lot of irrelevant information.
Over the years, C-Reduce has evolved into a sophisticated hybrid input reduction tool, utilizing the following transformations.
\begin {itemize}
\item Delta debugging using topformflat~\cite{delta}
\item Various text- and tree-based transformations over the source code, for example: function inlining, removal of unused functions and variables, etc.
\item Source code pretty printing
\end {itemize}
C-Reduce has a proven track record of being able to handle even complex C/C\texttt{++} programs~\cite{groce2016cause}.
ReduKtor may be considered an adaptation of the C-Reduce approach to the Kotlin programming language; however, it also supports simultaneous reduction of several Kotlin files, i.e., project-level input reduction, and uses slicing to improve performance.

Other tools for input reduction of Java or C\texttt{++} programming languages include JSlice~\cite{WR:04}, Indus~\cite{jayaraman2005kaveri}, JavaBST~\cite{abdallah2017javabst} and CodeSurfer~\cite{anderson2004codesurfer}.
They implement a variety of techniques, in a way more or less similar to C-Reduce.
An example of an approach, similar to ours, combining delta debugging and static slicing would be the one by Leitner et al.~\cite{leitner2007efficient}, targeted at minimizing randomly generated tests for Eiffel.

Some approaches use dynamic slicing instead of static, which helps with the related activity of program understanding, e.g., Gupta et al.~\cite{gupta2005locating}.
This approach takes a forward dynamic slice from the delta debugged input and intersects it with a backward dynamic slice of the erroneous output, creating what is called a failure-inducing chop.
It is used to better guide the bug-finding activities; we did not have the opportunity to focus on this problem in our work, and hope to explore it in the future.

A number of tools attempt to perform language-agnostic input reduction.
The classic example of this is Picireny~\cite{hodovan2017tree}, an implementation of~HDD.
It performs HDD over the parse trees produced from an input ANTLR~\cite{parr2013definitive} grammar; if your input can be described with ANTLR, Picireny can try to reduce it.
The advantages and disadvantages of this tool mirror the ones of HDD: it is universal, but often takes very long to reduce complex inputs.
Other tools based on HDD include~\cite{lei2005minimization, orso2006isolating}.

Herfert in~\cite{herfert2017automatically} presents a language-independent algorithm, named Generalized Tree Reduction~(GTR).
It extends HDD, in an attempt to improve its performance, combining HDD with a greedy backtracking-based search over a set of generic tree transformations.
These transformations improve the HDD performance and efficiency, by limiting the number of tree configurations considered on every level and also allowing to reduce the tree on several levels simultaneously.

Sun et al.~\cite{sun2018perses} tackle the same problem~(HDD performance) from another angle in their Perses framework.
Perses is also language-agnostic and accepts input grammar in Backus-Naur form, which is then transformed into an internal grammar representation.
Tree nodes are classified into four classes (regular, Kleene-Plus, Kleene-Star and Optional); regular node is replaced with its minimal compatible descendant, other types of nodes are reduced using a variation of delta debugging.
Evaluation shows Perses produces much smaller results compared to most other input reduction tools~(55-98\% smaller) except for C-Reduce.
At the same time, it performs on average two times faster than C-Reduce.
Gharachorlu et al.~\cite{gharachorlu2019priority} present a new technique called Pardis, which improves Perses by prioritizing reduction of larger subtrees first.
It is shown to work 1.3x to 7.8x faster than Perses, with less reduction soundness checks and same overall reduction quality.

\section{Conclusion}
\label{sec:conclusion}

Despite all the latest advances in software engineering, input reduction is still a very hard problem, in a lot of cases solved by tedious manual efforts.
The more complex an input structure is, the more time it takes to reduce the input manually; this is especially true for compilers, as reducing code requires deep understanding of the said code.

In this experience paper we present an approach to automatic input reduction for the Kotlin programming language.
The approach is based on a combination of Kotlin-specific transformations, program slicing, and hierarchical delta debugging, which are highly synergistic.
It also accounts for multi-file projects by supporting simultaneous reduction of several  files.

We have implemented a prototype tool called ReduKtor based on our approach, and performed its thorough evaluation.
The results show that, to achieve high reduction quality, one must still employ language-specific transformations together with general approaches, such as~HDD; language-agnostic techniques, despite their recent advances, still fail to achieve the efficiency of custom transformations when used standalone.

As for our future work, we hope to explore both theoretical and practical improvements.
The input reduction pipeline consists of multiple steps, and the order of these steps may influence the performance; it would be interesting to see if machine learning could be used to find a quazi-optimal ordering, depending on the input properties.
From the practical side, we plan to better parallelize the pipeline, allowing to perform transformations simultaneously, and improve the integration with Kotlin compiler, to speed up the soundness checks.
We also hope to collaborate with the Kotlin compiler team on incorporating ReduKtor into their workflow, to further the understanding of how input reduction performs in practice.

\section*{Acknowledgement}

We would like to express our gratitude to the Kotlin compiler team for their support and feedback, to all the reviewers for their questions and comments, and to our shepherd for their help in improving the final version of the paper.

\balance

\bibliographystyle{IEEEtran}
\bibliography{IEEEabrv,reduktor}

\begin{thebibliography}{10}
\providecommand{\url}[1]{#1}
\csname url@samestyle\endcsname
\providecommand{\newblock}{\relax}
\providecommand{\bibinfo}[2]{#2}
\providecommand{\BIBentrySTDinterwordspacing}{\spaceskip=0pt\relax}
\providecommand{\BIBentryALTinterwordstretchfactor}{4}
\providecommand{\BIBentryALTinterwordspacing}{\spaceskip=\fontdimen2\font plus
\BIBentryALTinterwordstretchfactor\fontdimen3\font minus
  \fontdimen4\font\relax}
\providecommand{\BIBforeignlanguage}[2]{{%
\expandafter\ifx\csname l@#1\endcsname\relax
\typeout{** WARNING: IEEEtran.bst: No hyphenation pattern has been}%
\typeout{** loaded for the language `#1'. Using the pattern for}%
\typeout{** the default language instead.}%
\else
\language=\csname l@#1\endcsname
\fi
#2}}
\providecommand{\BIBdecl}{\relax}
\BIBdecl

\bibitem{zeller2002simplifying}
A.~Zeller and R.~Hildebrandt, ``Simplifying and isolating failure-inducing
  input,'' \emph{IEEE Transactions on Software Engineering}, vol.~28, no.~2,
  pp. 183--200, 2002.

\bibitem{weiser1981program}
M.~Weiser, ``Program slicing,'' in \emph{International Conference on Software
  Engineering}.\hskip 1em plus 0.5em minus 0.4em\relax IEEE Press, 1981, pp.
  439--449.

\bibitem{misherghi2006hdd}
G.~Misherghi and Z.~Su, ``{HDD}: Hierarchical delta debugging,'' in
  \emph{International Conference on Software Engineering}.\hskip 1em plus 0.5em
  minus 0.4em\relax ACM, 2006, pp. 142--151.

\bibitem{herfert2017automatically}
S.~Herfert, J.~Patra, and M.~Pradel, ``Automatically reducing tree-structured
  test inputs,'' in \emph{International Conference on Automated Software
  Engineering}.\hskip 1em plus 0.5em minus 0.4em\relax IEEE, 2017, pp.
  861--871.

\bibitem{binkley2014orbs}
D.~Binkley, N.~Gold, M.~Harman, S.~Islam, J.~Krinke, and S.~Yoo, ``{ORBS}:
  Language-independent program slicing,'' in \emph{International Symposium on
  Foundations of Software Engineering}.\hskip 1em plus 0.5em minus 0.4em\relax
  ACM, 2014, pp. 109--120.

\bibitem{sterling2007automated}
C.~D. Sterling and R.~A. Olsson, ``Automated bug isolation via program
  chipping,'' \emph{Software: Practice and Experience}, vol.~37, no.~10, pp.
  1061--1086, 2007.

\bibitem{regehr2012test}
J.~Regehr, Y.~Chen, P.~Cuoq, E.~Eide, C.~Ellison, and X.~Yang, ``Test-case
  reduction for {C} compiler bugs,'' in \emph{ACM SIGPLAN Conference on
  Programming Language Design and Implementation}, vol.~47, no.~6.\hskip 1em
  plus 0.5em minus 0.4em\relax ACM, 2012, pp. 335--346.

\bibitem{fuzzer2017}
M.~Koltsov, ``Kotlin fuzzer,''
  \url{https://github.com/ItsLastDay/KotlinFuzzer}, 2017, accessed: 01.05.2019.

\bibitem{jsDelta}
``{JS Delta},'' \url{https://github.com/wala/jsdelta}, 2019, accessed:
  01.05.2019.

\bibitem{myers1986ano}
E.~W. Myers, ``An {O(ND)} difference algorithm and its variations,''
  \emph{Algorithmica}, vol.~1, no. 1-4, pp. 251--266, 1986.

\bibitem{diff-match-patch}
``Diff match patch library,'' \url{https://github.com/google/diff-match-patch},
  2019, accessed: 01.05.2019.

\bibitem{2002slicing}
V.~Kasyanov and I.~Mirzuitova, ``Slicing: Program slices and their
  applications,'' p. 116, 2002.

\bibitem{WR:04}
T.~Wang and A.~Roychoudhury, ``Using compressed bytecode traces for slicing
  {Java} programs,'' in \emph{International Conference on Software
  Engineering}, 2004, pp. 512--521.

\bibitem{gharachorlu2019priority}
G.~Gharachorlu and N.~Sumner, ``Pardis: Priority aware test case reduction,''
  in \emph{International Conference on Fundamental Approaches to Software
  Engineering}.\hskip 1em plus 0.5em minus 0.4em\relax Springer, 2019, pp.
  409--426.

\bibitem{yang2011finding}
X.~Yang, Y.~Chen, E.~Eide, and J.~Regehr, ``Finding and understanding bugs in
  {C} compilers,'' in \emph{ACM SIGPLAN Conference on Programming Language
  Design and Implementation}, vol.~46, no.~6.\hskip 1em plus 0.5em minus
  0.4em\relax ACM, 2011, pp. 283--294.

\bibitem{delta}
S.~McPeak and D.~S. Wilkerson, ``Delta debugging implementation,'' \url{
  http://delta.tigris.org/}, accessed: 01.05.2019.

\bibitem{groce2016cause}
A.~Groce, M.~A. Alipour, C.~Zhang, Y.~Chen, and J.~Regehr, ``Cause reduction:
  Delta debugging, even without bugs,'' \emph{Software Testing, Verification
  and Reliability}, vol.~26, no.~1, pp. 40--68, 2016.

\bibitem{jayaraman2005kaveri}
G.~Jayaraman, V.~P. Ranganath, and J.~Hatcliff, ``Kaveri: Delivering the
  {Indus} {Java} program slicer to {Eclipse},'' in \emph{International
  Conference on Fundamental Approaches to Software Engineering}.\hskip 1em plus
  0.5em minus 0.4em\relax Springer, 2005, pp. 269--272.

\bibitem{abdallah2017javabst}
M.~Abdallah, B.~Alokush, M.~Alrefaee, M.~Salah, R.~Bader, and K.~Awad,
  ``{JavaBST}: Java backward slicing tool,'' in \emph{International Conference
  on Information Technology}.\hskip 1em plus 0.5em minus 0.4em\relax IEEE,
  2017, pp. 614--618.

\bibitem{anderson2004codesurfer}
P.~Anderson, ``{CodeSurfer/Path} inspector,'' in \emph{International Conference
  on Software Maintenance}.\hskip 1em plus 0.5em minus 0.4em\relax IEEE, 2004,
  p. 508.

\bibitem{leitner2007efficient}
A.~Leitner, M.~Oriol, A.~Zeller, I.~Ciupa, and B.~Meyer, ``Efficient unit test
  case minimization,'' in \emph{International Conference on Automated Software
  Engineering}.\hskip 1em plus 0.5em minus 0.4em\relax ACM, 2007, pp. 417--420.

\bibitem{gupta2005locating}
N.~Gupta, H.~He, X.~Zhang, and R.~Gupta, ``Locating faulty code using
  failure-inducing chops,'' in \emph{International Conference on Automated
  Software Engineering}.\hskip 1em plus 0.5em minus 0.4em\relax ACM, 2005, pp.
  263--272.

\bibitem{hodovan2017tree}
R.~Hodov{\'a}n, {\'A}.~Kiss, and T.~Gyim{\'o}thy, ``Tree preprocessing and test
  outcome caching for efficient hierarchical delta debugging,'' in
  \emph{International Workshop on Automation of Software Testing}.\hskip 1em
  plus 0.5em minus 0.4em\relax IEEE, 2017, pp. 23--29.

\bibitem{parr2013definitive}
T.~Parr, \emph{The definitive {ANTLR} 4 reference}.\hskip 1em plus 0.5em minus
  0.4em\relax Pragmatic Bookshelf, 2013.

\bibitem{lei2005minimization}
Y.~Lei and J.~H. Andrews, ``Minimization of randomized unit test cases,'' in
  \emph{International Symposium on Software Reliability Engineering}.\hskip 1em
  plus 0.5em minus 0.4em\relax IEEE, 2005, pp. 267--276.

\bibitem{orso2006isolating}
A.~Orso, S.~Joshi, M.~Burger, and A.~Zeller, ``Isolating relevant component
  interactions with {JINSI},'' in \emph{International Workshop on Dynamic
  Systems Analysis}.\hskip 1em plus 0.5em minus 0.4em\relax ACM, 2006, pp.
  3--10.

\bibitem{sun2018perses}
C.~Sun, Y.~Li, Q.~Zhang, T.~Gu, and Z.~Su, ``Perses: Syntax-guided program
  reduction,'' in \emph{International Conference on Software
  Engineering}.\hskip 1em plus 0.5em minus 0.4em\relax ACM, 2018, pp. 361--371.

\end{thebibliography}

\end{document}